\documentclass[12pt]{article}

\usepackage{cite}
\usepackage{epsfig}
\usepackage{amsmath}
\usepackage{array}
\usepackage{pstricks}
\usepackage{bbold}

\def\mathswitch#1{\relax\ifmmode#1\else$#1$\fi}
\def\mathswitchr#1{\relax\ifmmode{\mathrm{#1}}\else$\mathrm{#1}$\fi}
\newcommand{\PW}{\mathswitchr W}
\newcommand{\PZ}{\mathswitchr Z}
\newcommand{\PH}{\mathswitchr H}

\newcommand{\Pb}{\mathswitchr b}
\newcommand{\Pt}{\mathswitchr t}

\newcommand{\MW}{\mathswitch {M_\PW}}
\newcommand{\MZ}{\mathswitch {M_\PZ}}
\newcommand{\MH}{\mathswitch {M_\PH}}

\newcommand{\mb}{\mathswitch {m_\Pb}}
\newcommand{\mt}{\mathswitch {m_\Pt}}

\newcommand{\mw}{\mathswitch {\overline{M}_\PW}}
\newcommand{\mz}{\mathswitch {\overline{M}_\PZ}}

\newcommand{\as}{\alpha_{\mathrm s}}

\newcommand{\tev}{\,\, \mathrm{TeV}}
\newcommand{\gev}{\,\, \mathrm{GeV}}

\newcommand{\re}{\Re e \,}

\newcommand{\OO}{{\mathcal O}}

\newcommand{\mycaption}[1]{\caption{\sl #1}}

\begin{document}
\thispagestyle{empty}

\def\thefootnote{\fnsymbol{footnote}}

\begin{flushright}
\end{flushright}

\vspace{1cm}

\begin{center}

{\Large\sc {\bf Electroweak two-loop corrections to \boldmath 
$\sin^2 \theta_{\rm eff}^{b\bar{b}}$ and $R_b$
\\[.5ex]
 using numerical Mellin-Barnes integrals}}
\\[3.5em]
{\large\sc
Ayres~Freitas, Yi-Cheng~Huang
}

\vspace*{1cm}

{\sl
Pittsburgh Particle-physics Astro-physics \& Cosmology Center
(PITT-PACC),\\ Department of Physics \& Astronomy, University of Pittsburgh,
Pittsburgh, PA 15260, USA
}

\end{center}

\vspace*{2.5cm}

\begin{abstract}
Multi-loop integrals can be evaluated numerically using Mellin-Barnes
representations. Here this technique is applied to the calculation of
electroweak two-loop correction with closed fermion loops for two observables:
the effective weak mixing angle for bottom quarks, $\sin^2 \theta_{\rm
eff}^{b\bar{b}}$, and the branching ratio of the $Z$ boson into bottom quarks,
$R_b$. Good agreement with a previous result for $\sin^2 \theta_{\rm
eff}^{b\bar{b}}$ is found. The result for $R_b$ is new, and a 
simple parametrization formula is
provided which approximates the full result within integration errors.
\end{abstract}

\setcounter{page}{0}
\setcounter{footnote}{0}

\newpage


\section{Introduction}

High-precision data from LEP, SLC, the Tevatron, and the LHC allow us to perform
very accurate tests of the Standard Model (SM), predict the mass of the Higgs
boson, and possibly identify hints for new physics. For this purpose, the
experimental results are compared to theoretical computations including
higher-order radiative corrections. For some of the most precisely determined
observables it is necessary to include complete two-loop corrections in the
calculation. Such calculations have been carried out for the mass of the $W$
boson, $\MW$ \cite{mw,mwlong,mwtot}, and the effective weak mixing angle $\sin^2
\theta_{\rm eff}$ parametrizing the ratio of vector and axial-vector couplings
of the $Z$ boson to leptons \cite{swlept,swlept2,swlept3}, light quarks
\cite{swlept2}, and bottom quarks \cite{swbb}. Furthermore, partial electroweak
three-loop contributions \cite{mt6}, as well as three-loop \cite{qcd3,qcd3light}
and leading four-loop \cite{qcd4} QCD corrections to these observables have been
determined. The two-loop and leading three-loop results for $\MW$ and $\sin^2
\theta_{\rm eff}$ have been implemented in commonly used SM fit programs such as
\textsc{ZFitter} \cite{zfitter} and \textsc{GFitter} \cite{gfitter}.

A major difficulty in these calculations are two-loop integrals
with multiple mass and momentum scales. In general these
integrals cannot be solved analytically in closed form, so that one has to use
numerical methods instead. Numerical techniques for the evaluation of loop 
integrals face two main challenges: extraction of ultraviolet (UV)
and infrared (IR) singularities, as well as stability and efficiency of the
numerical integrations. A powerful method is based on Mellin-Barnes
(MB) representations. It has been demonstrated that MB representations can be
used  to isolate the singularities of an arbitrary multi-loop integral in a
systematic way \cite{mb,mb2}, such that this procedure can be automated in a
computer program \cite{mb2,ambre}. Furthermore, it was shown that the
integration time and convergence behavior of the MB integrals can be improved
substantially by suitable variable transformations and by analytically
integrating over some variables of the multi-dimensional MB integrals
\cite{mbn}.

This article reports on the concrete calculation of two electroweak precision
observables using the technique described in Ref.~\cite{mbn}. First, the method
is validated by reproducing the result for the fermionic electroweak  two-loop
corrections to the effective weak mixing angle of bottom quarks, $\sin^2
\theta_{\rm eff}^{b\bar{b}}$, published earlier in Ref.~\cite{swbb}. Here the
term ``fermionic'' refers to diagrams with at least one closed fermion loop.
Secondly, a new result is presented for the complete fermionic electroweak
two-loop (next-to-next-leading order) corrections to $R_b$, the branching ratio
of the $Z$ boson into bottom quarks and all quarks. Until now, only an
approximate results for the electroweak two-loop corrections to the $Z$ partial
widths are available, using an expansion for large values of the top-quark mass.
For decays of the $Z$ boson into quarks of the first two generations, this
expansion has been driven to the order $\OO(\alpha^2\mt^2)$ \cite{ewmt2}, while
for the $Z\to b\bar{b}$ width only the leading $\OO(\alpha^2\mt^4)$ coefficient
is known \cite{ewmt4}.

As elaborated in the next section, the radiative corrections for both $\sin^2
\theta_{\rm eff}^{b\bar{b}}$ and $R_b$ are obtained from the same loop diagrams
of the $Zf\bar{f}$ vertex. For the calculation presented here, the MB method of
Ref.~\cite{mbn} has been used for the most difficult two-loop diagrams involving
triangle sub-loops. The remaining diagrams have been computed by reducing the
relevant integrals to a set of master integrals \cite{ibp}, which are then
evaluated numerically \cite{bauberger,swlept2}. These steps are described in more
detail in section \ref{sc:2l}. Finally, numerical results for both observables
are presented in section~\ref{sc:res}, and the main findings are summarized in
section~\ref{sc:sum}.


\section{\boldmath $\sin^2 \theta_{\rm eff}^{b\bar{b}}$ and $R_b$ at
next-to-next-to-leading order}
\label{sc:def}

\begin{figure}[tb]
\begin{center}
\raisebox{3cm}{(a)}\psfig{figure=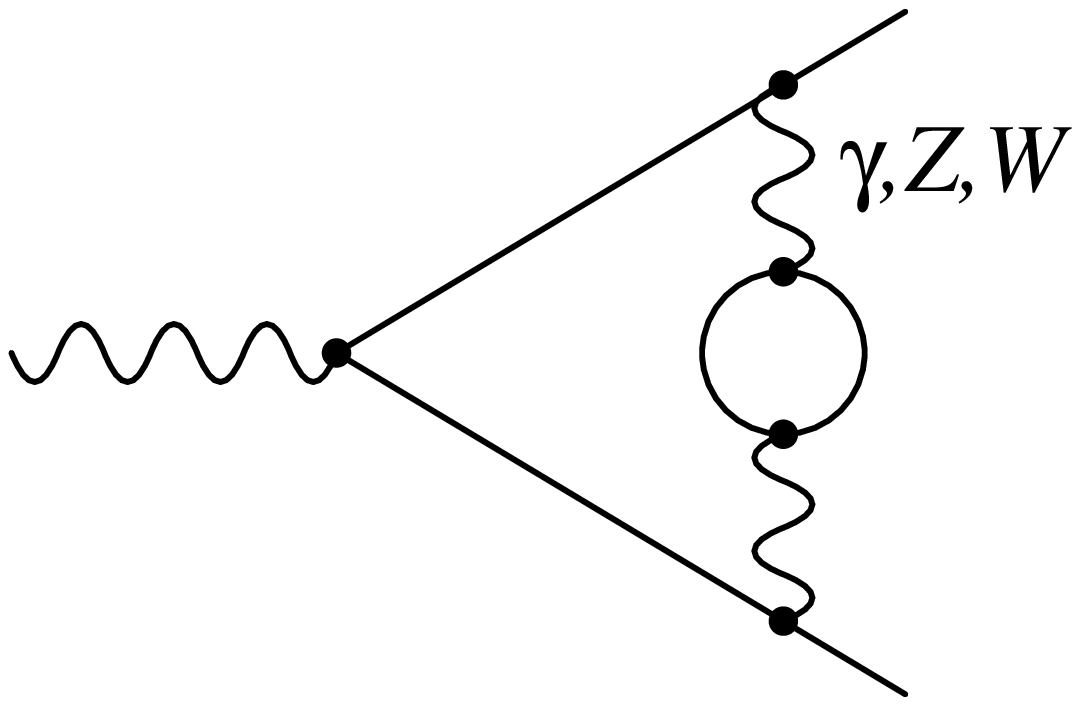, width=5cm}
\raisebox{3cm}{(b)}\psfig{figure=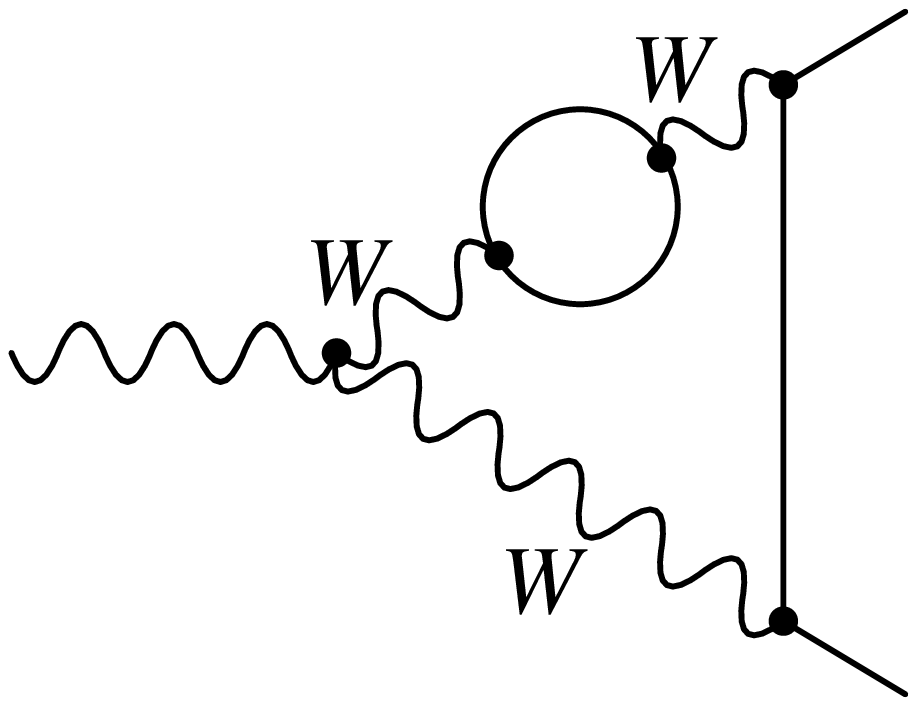, width=3.9cm}\\[1ex]
\raisebox{3cm}{(c)}\psfig{figure=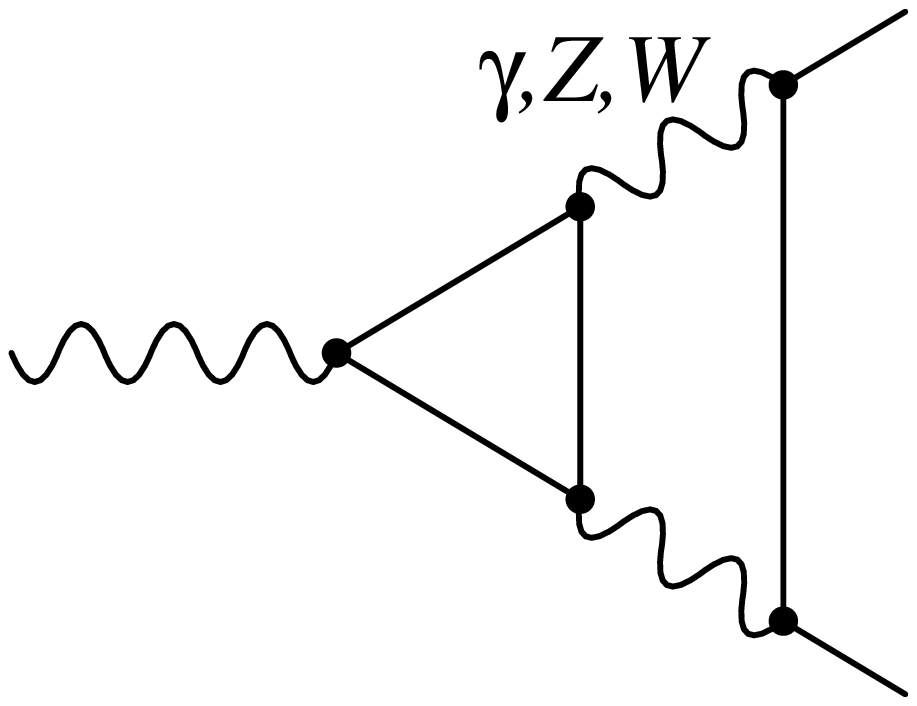, width=3.9cm} \hspace{10mm}
\raisebox{3cm}{(d)}\psfig{figure=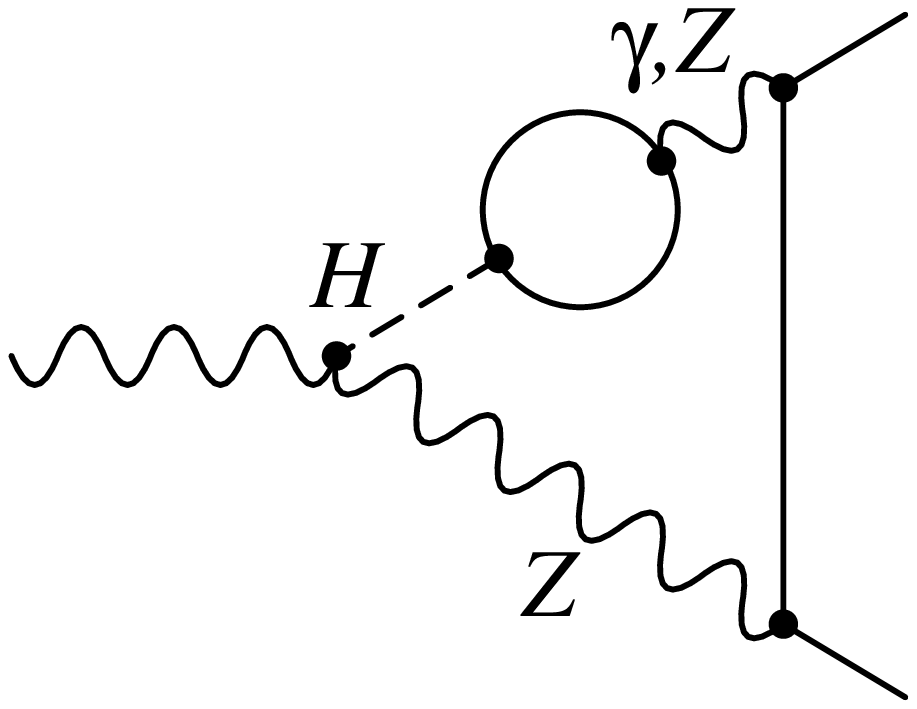, width=3.9cm}
\end{center}
\mycaption{Fermionic electroweak two-loop $Zf\bar{f}$ vertex diagrams contributing to 
the vertex form factors $v_f^{(2)}$ and $a_f^{(2)}$.
\label{fig:dia}}
\end{figure}

The effective weak mixing angle $\sin^2 \theta_{\rm eff}^{f\bar{f}}$ is related
to the ratio of vector and axial-vector couplings, $v_{\rm f}$ and $a_{\rm f}$
of the $Zf\bar{f}$ vertex, $i.\,e.$
\begin{equation}
\sin^2 \theta_{\rm eff}^{f\bar{f}} \equiv \frac{1}{4}\Bigl (
 1 + \re \frac{v_f}{a_f} \Bigr ).
\end{equation}
Expanding this definition to next-to-next-to-leading order (NNLO) one obtains
\begin{equation}
\begin{aligned}
\sin^2 \theta_{\rm eff}^{f\bar{f}} &= \left(1- \frac{\mw^2}{\mz^2} \right) \re\Biggl\{ 1+
   \frac{a_f^{(1)} \, v_f^{(0)} -
        v_f^{(1)} \, a_f^{(0)}}{a_f^{(0)}(a_f^{(0)}-v_f^{(0)})}
        \Biggr|_{k^2 = \MZ^2} \\
 &\hspace{6em}
     + \frac{a_f^{(2)} \, v_f^{(0)} \, a_f^{(0)} -
        v_f^{(2)} \, (a_f^{(0)})^2 -
        (a_f^{(1)})^2 \, v_f^{(0)} +
        a_f^{(1)} \, v_f^{(1)} \, a_f^{(0)}}{(a_f^{(0)})^2
        (a_f^{(0)}-v_f^{(0)})} \Biggr|_{k^2 = \MZ^2}
   \Biggr\},
\end{aligned} \label{eq:sin2}
\end{equation}
where $v_f^{(0,1,2)}$ denote the tree-level, one-loop, and two-loop
contributions to the vector form factor, respectively, and $a_f^{(0,1,2)}$ are
defined similarly for the axial-vector form factor. Figure~\ref{fig:dia} shows
the types of two-loop diagrams with closed fermions loops that contribute to 
$v_f^{(2)}$ and $a_f^{(2)}$.
$\mw^2$ and $\mz^2$ are the real parts of the gauge-invariant
propagator poles of the $W$ and $Z$ boson, respectively.
In eq.~\eqref{eq:sin2} the infrared (IR) divergent QED and QCD contributions cancel exactly,
see for example Ref.~\cite{swlept2}, so that it is sufficient to only include
IR-finite weak loop corrections.

\vspace{\medskipamount}
The branching ratio $R_b$ is defined as the ratio of the partial decay widths of
the $Z$-boson decay into bottom quarks and into all quarks:
\begin{equation}
R_b \equiv \frac{\Gamma_b}{\Gamma_{\rm had}}
 = \frac{\Gamma_b}{\Gamma_d+\Gamma_u+\Gamma_s+\Gamma_c+\Gamma_b}
 = \frac{1}{1+2(\Gamma_d+\Gamma_u)/\Gamma_b}, \label{eq:rb}
\end{equation}
where $\Gamma_f$ stands for the partial decay width into the $f\bar{f}$ final
state. In the last step in \eqref{eq:rb}, the relationships $\Gamma_u \approx
\Gamma_c$ and $\Gamma_d \approx \Gamma_s$ have been used, which hold to very good
approximation.

Up to next-to-next-to-leading order ($q=u,d$), 
\begin{align}
\frac{\Gamma_q}{\Gamma_b} = \frac{G_q^{(0)}}{G_b^{(0)}}
\;+\; &\frac{2}{(G_b^{(0)})^2} \,\re\! \Bigl\{
   G_b^{(0)}G_q^{(1)} - G_q^{(0)}G_b^{(1)}
  \Bigr \} + \frac{1}{(G_b^{(0)})^2} \bigl [ 
   G_b^{(0)} {\cal R}_q^{(1)} - G_q^{(0)} {\cal R}_b^{(1)} \bigr ] \nonumber \\
\;+\; &\frac{1}{(G_b^{(0)})^3} \,\re\! \Bigl\{
 \begin{aligned}[t]
 & (G_b^{(0)})^2 \bigl [2G_q^{(2)} + (a_q^{(1)})^2 + (v_q^{(1)})^2 \bigr ] \nonumber \\
 &- G_b^{(0)} G_q^{(0)} \bigl [2G_b^{(2)} + (a_b^{(1)})^2 + (v_b^{(1)})^2 \bigr ] \\
 &- 4G_b^{(0)} G_q^{(1)} G_b^{(1)} 
  + 4G_q^{(0)} (G_b^{(1)})^2 \Bigr\}
 \end{aligned} \\
 &+ \frac{1}{(G_b^{(0)})^2} \bigl [ 
   G_b^{(0)} {\cal R}_q^{(2)} -
   G_q^{(0)} {\cal R}_b^{(2)} -
   G_b^{(0)} {\cal R}_q^{(1)} {\cal R}_b^{(1)} +
   G_q^{(0)} ({\cal R}_b^{(1)})^2 \bigr ] \label{eq:rbe} \\
 &+ \frac{2}{(G_b^{(0)})^3} \bigl [
   (G_b^{(0)})^2 (a_q^{(0)} a_q^{(1)} {\cal R}_{q,\rm A}^{(1)} +
                   v_q^{(0)} v_q^{(1)} {\cal R}_{q,\rm V}^{(1)})
   -G_b^{(0)} G_b^{(1)} {\cal R}_q^{(1)} \nonumber \\
 &\quad
   +G_q^{(0)} G_b^{(1)} {\cal R}_b^{(1)}
   +G_q^{(0)} a_b^{(0)}v_b^{(0)} (a_b^{(0)}v_b^{(1)} + v_b^{(0)}a_b^{(1)})
    ({\cal R}_{b,\rm A}^{(1)} - {\cal R}_{b,\rm V}^{(1)}) \nonumber \\
 &\quad
   -G_b^{(0)} G_q^{(1)}  {\cal R}_b^{(1)}
 \bigr ], \nonumber 
\end{align}
with
\begin{equation}
G_q^{(n)} = a_q^{(0)}a_q^{(n)} + v_q^{(0)}v_q^{(n)}, \qquad
\qquad 
{\cal R}_q^{(n)} = (a_q^{(0)})^2 {\cal R}_{q,\rm A}^{(n)} +
	(v_q^{(0)})^2 {\cal R}_{q,\rm V}^{(n)}	.
\end{equation}
Here ${\cal R}_{q,\rm V}^{(n)}$ and ${\cal R}_{q,\rm A}^{(n)}$ incorporate the
$n$-loop QED and QCD corrections to the vector and axial-vector form factors,
which have been calculated already several years ago \cite{RVA, RVA2}, see also
Ref.~\cite{RVA3}. The relevant parts for this calculation are given by
\begin{align} 
{\cal R}_{d,\rm V}^{(1)} &= {\cal R}_{d,\rm A}^{(1)} = 
 {\cal R}_{b,\rm V}^{(1)} = \tfrac{\alpha}{12\pi} + \tfrac{\as}{\pi}, \\ 
{\cal R}_{u,\rm V}^{(1)} &= {\cal R}_{u,\rm A}^{(1)} = 
 \tfrac{\alpha}{3\pi} + \tfrac{\as}{\pi}, \\ 
{\cal R}_{b,\rm A}^{(1)} &= {\cal R}_{b,\rm V}^{(1)} - 6\tfrac{\mb^2}{\MZ^2},
\label{eq:rba1} \\[1ex] 
{\cal R}_{d,\rm V}^{(2)} &= -\tfrac{\alpha\as}{36\pi^2} + C_2 
 \bigl(\tfrac{\as}{\pi}\bigr)^2 + C_3 \bigl(\tfrac{\as}{\pi}\bigr)^3, \\ 
{\cal R}_{d,\rm A}^{(2)} &= {\cal R}_{d,\rm V}^{(2)}
 + I_2\bigl(\tfrac{\MZ^2}{\mt^2}\bigr)\, \bigl(\tfrac{\as}{\pi}\bigr)^2
 + I_3\bigl(\tfrac{\MZ^2}{\mt^2}\bigr)\, \bigl(\tfrac{\as}{\pi}\bigr)^3, \\ 
{\cal R}_{u,\rm V}^{(2)} &= {\cal R}_{d,\rm V}^{(2)} -\tfrac{\alpha\as}{12\pi^2}, \\ 
{\cal R}_{u,\rm A}^{(2)} &= {\cal R}_{u,\rm V}^{(2)}
 - I_2\bigl(\tfrac{\MZ^2}{\mt^2}\bigr)\, \bigl(\tfrac{\as}{\pi}\bigr)^2
 - I_3\bigl(\tfrac{\MZ^2}{\mt^2}\bigr)\, \bigl(\tfrac{\as}{\pi}\bigr)^3,\\ 
{\cal R}_{b,\rm V}^{(2)} &= {\cal R}_{d,\rm V}^{(2)} + 12\tfrac{\mb^2}{\MZ^2}
 \tfrac{\as}{\pi} - 6\tfrac{\mb^4}{\MZ^4} + \OO(\mb^4\as,\, \mb^2\as^2),  \label{eq:rbv2} \\ 
{\cal R}_{b,\rm A}^{(2)} &= {\cal R}_{d,\rm A}^{(2)} - 22\tfrac{\mb^2}{\MZ^2}
 \tfrac{\as}{\pi} + 6\tfrac{\mb^4}{\MZ^4} + \OO(\mb^4\as,\, \mb^2\as^2).
 \label{eq:rba2}
\end{align}
Here three-loop QCD corrections are also included in the ${\cal R}^{(2)}$ terms,
and $\mb$ is defined in the $\overline{\rm MS}$-scheme. In the
power-counting the small bottom quark mass is treated to be parametrically of
the same order as the fine-structure constant, $\tfrac{\mb^2}{\MZ^2} \sim
\alpha$, so that only the leading tree-level $\mb$-dependence is included in
\eqref{eq:rba1} and the leading one-loop $\mb$-dependence in
\eqref{eq:rbv2},~\eqref{eq:rba2}. Note that
the dependence on $C_2$ and $C_3$ drops out in the final result for $R_b$.
The functions $I_2$ and $I_3$ read
\begin{align}
I_2(x) &= -\frac{37}{12} + \log x + \frac{7}{81} x + 0.0132\, x^2, \\[.5ex]
I_3(x) &= -\frac{5075}{216} + \frac{23}{6}\,\zeta_2 + \zeta_3 +
  \frac{67}{18}\log x + \frac{23}{12}\log^2 x.
\end{align}
The vertex form factors $v_f^{(2)}$ and $a_f^{(2)}$ include all electroweak
two-loop diagrams, as shown in Fig.~\ref{fig:dia}, except those that involve
IR-divergent photon exchange contributions between the outgoing fermion lines
(represented by Fig.~\ref{fig:dia}~(a) when both wiggly lines are photons).
For the effective weak mixing angle $\sin^2
\theta_{\rm eff}$ these QED diagrams cancel in the final result, 
as pointed out in Ref.~\cite{swlept2}.
For the branching fraction $R_b$, on the other hand, the QED contributions are
included in the radiator functions ${\cal R}_{q,\rm V}$ and ${\cal R}_{q,\rm
A}$. In consequence, for both observables, the IR-divergent QED diagrams are
excluded from the calculation of the electroweak two-loop corrections discussed
here.

\vspace{\medskipamount}
For the renormalization the on-shell scheme is employed. In particular,  the
renormalized squared masses are defined as the real part of the propagator poles
(which is different than the pole of the real part of the propagator).
Furthermore, the external fields are renormalized to unity at the position of
the poles. Details and explicit expressions for the renormalization constants
can be found in Ref.~\cite{mwlong}.

For the computation of the electroweak two-loop corrections, the masses and
Yukawa couplings of all fermions except the top quark can be safely neglected.
Moreover, the CKM matrix is approximated by the unit matrix.


\section{Outline of the computation of two-loop diagrams}
\label{sc:2l}

Let us start by giving a brief overview to the use of MB
representations for the numerical evaluation of multi-loop integrals. For more
details, see Refs.~\cite{mb,mb2,mbn}.

After introduction of Feynman parameters, a general one-loop integral with $N$
propagators can be written in the form
\begin{equation}
I = (-1)^N \Gamma(N-D/2)
\int_0^1 dx_1\cdots dx_N \; \frac{\delta(1-x_1 - \ldots -x_N)}
{\left[\sum_{i,j=1}^N K_{ij}x_i x_j + \sum_{i=1}^N L_i x_i + M - i\epsilon
 \right]^{N-D/2}},
\end{equation}
where $D$ is the number of space-time dimensions, and $K_{ij}$, $L_i$, and $M$
depend on the masses and external momenta of the propagators.
Using the MB representation,
\begin{equation}
\begin{aligned}
\frac{1}{(A_0+\ldots+A_m)^Z} = \frac{1}{(2\pi i)^m} \int_{{\cal C}_1} dz_1
\cdots &\int_{{\cal C}_m} dz_m \; A_1^{z_1} \cdots A_m^{z_m}
A_0^{-Z-z_1-\ldots-z_m} \\
&\times \frac{\Gamma(-z_1) \cdots \Gamma(-z_m)\Gamma(Z+z_1+\ldots+z_m)}{\Gamma(
Z)},
\end{aligned}
\end{equation}
the integral $I$ can be transformed into a form that depends on the Feynman
parameters only in terms of exponentials $x_i^{z_i}$. The integration contours ${\cal C}_i$ for $z_i$ are straight lines parallel to
the imaginary axis chosen such that all arguments of the gamma functions have
positive real parts.
MB representations for
multi-loop integrals can be obtained recursively. After integration over the
Feynman parameters, the remaining MB integrals only depend on gamma functions of
the integration variables and on the external parameters. For example, the
``sunset'' two-loop diagram is given by the MB integral
\vspace{1ex}
\begin{align}
\rule{1.5cm}{0mm}
\psline(-1.5,0)(1.5,0)
\pscircle(0,0){0.8}
\psdot[dotscale=1.5](-0.8,0)
\psdot[dotscale=1.5](0.8,0)
\rput[t](0,0.65){m_1}
\rput[t](0,-0.9){m_3}
\rput[t](0,-0.1){m_2}
\rput[lb](-1.5,0.1){\displaystyle p\atop\longrightarrow}
\rule{1.6cm}{0mm}
= \frac{-1}{(2\pi i)^3} \int &dz_1dz_2dz_3 \; 
(m_1^2)^{-\varepsilon-z_1-z_2} (m_2^2)^{z_2} (m_3^2)^{1-\varepsilon+z_1-z_3}
(-p^2)^{z_3} 
\nonumber\\  &
\times \Gamma(-z_2) \Gamma(-z_3)\Gamma(1+z_1+z_2) \Gamma(z_3-z_1) \\[.5ex]
&\times \frac{\Gamma(1-\varepsilon-z_2)\Gamma(\varepsilon+z_1+z_2)
\Gamma(\varepsilon-1-z_1+z_3)}{\Gamma(2-\varepsilon+z_3)}, \nonumber
\label{ss}
\end{align}
with $\varepsilon=(4-D)/2$. All UV and IR singularities are now 
in the poles of the gamma functions and can be extracted in a systematic way, see
Refs.~\cite{mb,mb2}.

The remaining integrals in $z_i$ over the complex contours ${\cal C}_i$ can, in
principle, be carried out numerically. However, in practice, the integrand is
often highly oscillatory and the integral, while formally existent, may converge
too slowly.

In Ref.~\cite{mbn}, two effective methods for improving the convergence
behavior of the numerical $z_i$ integrals have been discussed. One 
modification consists in rotating the $z_i$-integration contours in the complex
plane, which can produce an additional exponentially damped term in the
integrand, so that the integral vanishes faster for $|z_i|\to\infty$. By
rotating the contours for all variables $z_i$ in parallel it is ensured that no
poles of the gamma functions cross the contour. Secondly, some integrations of
the multi-dimensional MB integral can be performed analytically with the help of
the convolution theorem for Mellin transforms. For typical two-loop vertex
diagrams needed for the calculation of the NNLO corrections to the $Zb\bar{b}$
interaction, one can treat about half of the integrations in this way, and only
the remaining half will be carried out numerically. With these improvements, it
was shown in Ref.~\cite{mbn} that the most complicated scalar integrals needed
for the NNLO $Zb\bar{b}$ corrections can be evaluated in a few hours on a
single-core computer.

\vspace{\medskipamount}
In this work, the MB method has been used for the most complicated integrals
stemming from the diagrams with triangle sub-loops, see Fig.~\ref{fig:dia}~(c). 
The remaining two-loop integrals have been calculated by first reducing them to
a set of master integrals using integration-by-parts and Lorentz identity
relations \cite{ibp}. Analytical formulas are known for the one-loop and
two-loop vacuum master integrals, while the two-loop selfenergy and vertex
master integrals have been evaluated numerically using dispersion relations as
described in Ref~\cite{bauberger,swlept2}.


\section{Results}
\label{sc:res}

The computational method described in the previous section has been applied to
the calculation of two-loop corrections to $\sin^2 \theta_{\rm eff}^{b\bar{b}}$
and $R_b$. Both quantities depend on the input parameters listed in the upper
part of Tab.~\ref{tab:input}. 
\begin{table}[tb]
\renewcommand{\arraystretch}{1.2}
\begin{center}
\begin{tabular}{ll}
\hline
Parameter & Experimental value \\
\hline
$\MZ$ & ($91.1876 \pm 0.0021$) GeV \\
$\Gamma_\PW$ & ($2.4952 \pm 0.0023$) GeV \\
$\MW$ & ($80.399 \pm 0.023$) GeV \\
$\Gamma_\PW$ & ($2.085 \pm 0.042$) GeV \\
$\mt$ & ($173.2 \pm 0.9$) GeV \\
$\Delta\alpha(\MZ)$ & $0.05900 \pm 0.00033$ \\
\hline
$\mb^{\overline{\rm MS}}$ & $4.20$ GeV \\
$\as(\MZ)$ & $0.1184 \pm 0.0007$ \\
$G_\mu$ & $1.16637 \times 10^{-5}$~GeV$^{-2}$\\
\hline
\end{tabular}
\end{center}
\vspace{-2ex}
\mycaption{Input parameters and their experimental values, from
Refs.~\cite{pdg,tevewwg}.
The electroweak two-loop corrections of the $Zb\bar{b}$ vertex depend only the first group of
parameters.
\label{tab:input}}
\end{table}
Note that the experimentally quoted values for the $W$- and $Z$-boson masses
correspond to a Breit-Wigner parametrization with a energy-dependent width, and
they have to be translation to the pole-mass scheme used in the loop calculation
\cite{masst}. In effect, this translation results in a downward shift of $\MW$ and
$\MZ$ by $\Gamma_\PW^2/(2\MW)$ and $\Gamma_\PZ^2/(2\MZ)$, respectively, where 
$\Gamma_{\PW,\PZ}$ and width of the gauge bosons.
$\Delta\alpha(\MZ)$ is the contribution from light
fermions (all fermions except the top quark) to the running of the
electromagnetic coupling between the scales $Q=0$ and $\MZ$.

\vspace{\medskipamount}

Let us first discuss of the effective
weak mixing angle of bottom quarks, $\sin^2 \theta_{\rm eff}^{b\bar{b}}$. The
fermionic two-loop corrections to this quantity have already been calculated
earlier \cite{swbb}, so that this result can be used as an additional check of
the MB method. The comparison is shown in Tab.~\ref{tab:swres} in terms of the
two-loop correction factor in the second line of eq.~\eqref{eq:sin2}:
\begin{table}[tb]
\renewcommand{\arraystretch}{1.2}
\begin{center}
\begin{tabular}{c@{\hspace{1.5em}}c@{\hspace{2em}}ll}
\hline
$\MW$ & $\mt$ & 
\multicolumn{2}{c}{$\Delta\kappa^{(\alpha^2,\mathrm{ferm})}_{b\overline{b}}
 \times 10^3$}
\\
\cline{3-4} 
[GeV] & [GeV] & Ref.~\cite{swbb} & this work \\ 
\hline
80.399 & 171.2 & $-2.278$ & $-2.303$ \\
80.399 & 172.2 & $-2.331$ & $-2.357$ \\
80.399 & 173.2 & $-2.386$ & $-2.410$ \\
80.399 & 174.2 & $-2.441$ & $-2.464$ \\
80.399 & 175.2 & $-2.496$ & $-2.518$ \\
80.422 & 173.2 & $-2.402$ & $-2.427$ \\
80.445 & 173.2 & $-2.418$ & $-2.444$ \\
\hline
\end{tabular}
\end{center}
\vspace{-2ex}
\mycaption{Results for the two-loop correction factor
$\Delta\kappa^{(\alpha^2,\mathrm{ferm})}_{b\overline{b}}$ to $\sin^2 \theta_{\rm
eff}^{b\bar{b}}$, for different values of $\MW$ and $\mt$. The other input
values have been set to $\MZ = 91.1876\gev$, $\MH = 100\gev$, $\Delta\alpha =
0$. The values obtained in this work are compared to the results of
Ref.~\cite{swbb}.
\label{tab:swres}}
\end{table}
\begin{equation}
\Delta\kappa^{(\alpha^2,\mathrm{ferm})}_{b\overline{b}}
 = \frac{a_b^{(2)} \, v_b^{(0)} \, a_b^{(0)} -
        v_b^{(2)} \, (a_b^{(0)})^2 -
        (a_b^{(1)})^2 \, v_b^{(0)} +
        a_b^{(1)} \, v_b^{(1)} \, a_b^{(0)}}{(a_b^{(0)})^2
        (a_b^{(0)}-v_b^{(0)})} \Biggr|_{k^2 = \MZ^2}
\end{equation}
The two calculations agree to within about 1\% of the two-loop contribution,
which is the same order as the uncertainty from the numerical MB integration.

\vspace{\medskipamount} 

Next, we turn to the presentation of results for the branching ratio $R_b$.
Table~\ref{tab:rbres1} lists the effects of various radiative correction terms,
starting with the one-loop contribution, corresponding to the first line of
eq.~\eqref{eq:rbe}, together with ${\cal O}(\as^2)$ final-state corrections,
which are numerically of similar order. 
The third column of Tab.~\ref{tab:rbres1} shows the new
result for the purely electroweak contributions from two-loop diagrams with
closed fermion loops. The fourth includes the electroweak two-loop corrections
together with final-state corrections of order ${\cal
O}(\as^3,\alpha\as,m_b^2\as,m_b^4)$. Finally, higher-order two-loop QCD corrections of order ${\cal
O}(\alpha\as)$ \cite{qcd2} to internal gauge-boson selfenergies, 
as well as three-loop QCD corrections of order
${\cal O}(\alpha\as^2)$ to the $\rho$-parameter \cite{qcd3} are given in the
last column of the table. Additional higher-order corrections to the
$\rho$-parameter \cite{qcd3light,qcd4,mt6,radja} are very small and have not
been included in the numerical analysis.

\begin{table}[tb]
\renewcommand{\arraystretch}{1.2}
\begin{center}
\begin{tabular}{c@{\hspace{1.5em}}c@{\hspace{1.5em}}c@{\hspace{1.5em}}c@{\hspace{1.5em}}c}
\hline
$\MH$ & ${\cal O}(\alpha) + {\rm FSR}_{\alpha,\as,\as^2}$ & 
${\cal O}(\alpha^2_{\rm ferm})$ &
${\cal O}(\alpha^2_{\rm ferm}) + {\rm FSR}_{\as^3,\alpha\as,m_b^2\as,m_b^4}$ &
${\cal O}(\alpha\as,\alpha\as^2)$
\\[-.5ex] 
[GeV] & [10$^{-4}$] & [10$^{-4}$] & [10$^{-4}$] & [10$^{-4}$] \\
\hline
\phantom{1}100 & $-$35.66 & $-$0.856 & $-$2.496 & $-$0.407 \\
\phantom{1}200 & $-$35.85 & $-$0.851 & $-$2.488 & $-$0.407 \\
\phantom{1}400 & $-$36.09 & $-$0.846 & $-$2.479 & $-$0.406 \\
\phantom{1}600 & $-$36.24 & $-$0.836 & $-$2.468 & $-$0.406 \\
1000           & $-$36.45 & $-$0.813 & $-$2.441 & $-$0.406 \\
\hline
\end{tabular}
\end{center}
\vspace{-2ex}
\mycaption{Results for electroweak one- and two-loop corrections to $R_b$, as defined in
eqs.~(\ref{eq:rb},\ref{eq:rbe}), for different values of $\MH$. The other input
values are taken from Tab.~\ref{tab:input}, with a fixed value for $\MW$. 
Also shown are the effects of two-
and three-loop QCD corrections to the final state (fourth column) and to 
gauge-boson selfenergies (fifth column). Here ``FSR'' stands for the final-state
radiative QCD and QED corrections described by the radiator functions ${\cal
R}^{(n)}$.
\label{tab:rbres1}}
\end{table}

As evident from the table, the electroweak two-loop corrections are relatively
small, which is a result of cancellations between individual contributions in
the result. 
The two-loop contribution depends only mildly on the Higgs-boson mass.
Its overall effect is a reduction of the SM prediction for $R_b$.
It is interesting to note that the electroweak two-loop
corrections, the final-state QED and QCD corrections, and the higher-order QCD 
contributions to internal gauge-boson lines all produce a
negative shift of the predicted values for $R_b$.
The uncertainty in the prediction for $R_b$
due to the numerical MB integration error is about $4\times 10^{-6}$.

\begin{figure}[tb]
\centering
\epsfig{figure=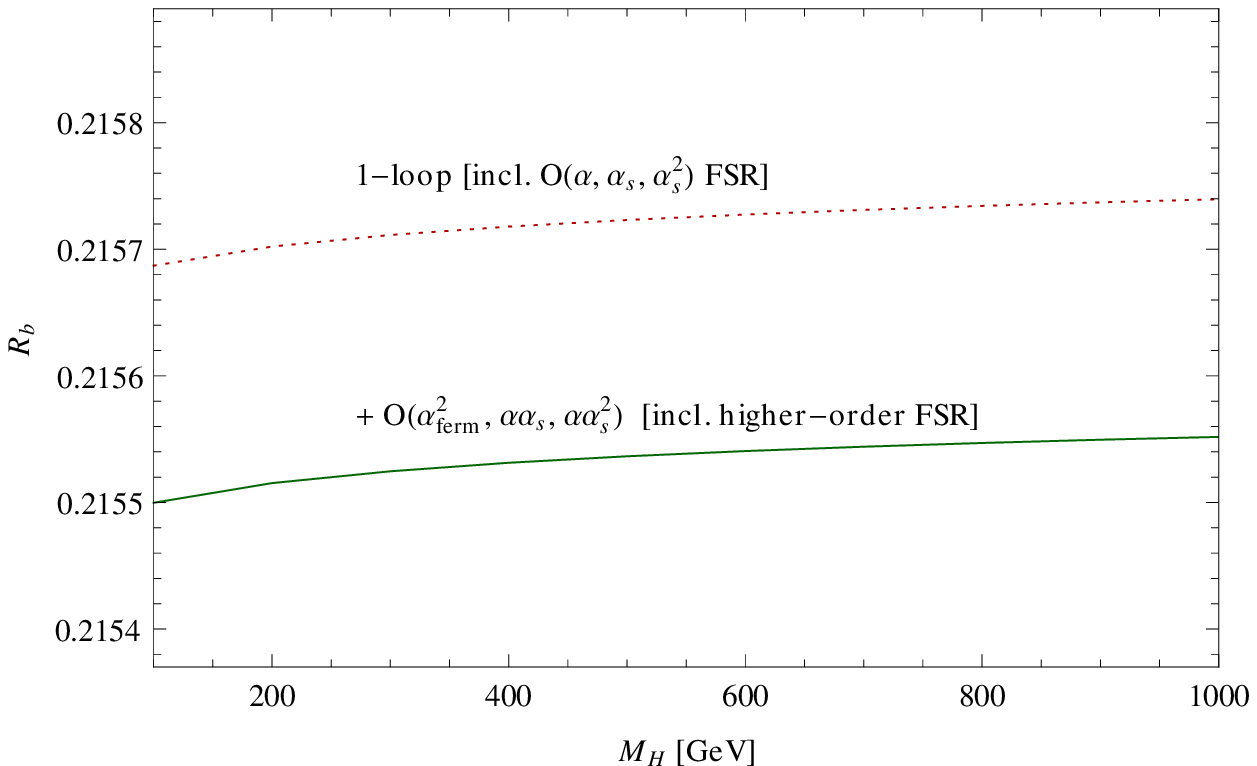, width=12cm}
\mycaption{One-loop and two-loop (with QCD three-loop contributions) result
for  the branching fraction $R_b$ as a function of $\MH$, and using the SM
prediction for $\MW$ to the same order of perturbation theory. Input values for
the other  parameters are taken from Tab.~\ref{tab:input}.
\label{fig:rbmh}}
\end{figure}

The numerical results in Tab.~\ref{tab:rbres1} have been
obtained with a fixed value of the $W$-boson mass. In global SM fits, however,
$\MW$ is calculated from the Fermi constant $G_\mu$. In
Refs.~\cite{mw,mwlong,mwtot} 
this calculation has been carried out including full two-loop and partial
higher-order corrections. Using the same order of perturbation theory for the
calculation of $\MW$ and $R_b$, the numerical results for $R_b$ in this scheme 
are given in Tab.~\ref{tab:rbres2} and Fig.~\ref{fig:rbmh}.
\begin{table}[tb]
\renewcommand{\arraystretch}{1.2}
\begin{center}
\begin{tabular}{c@{\hspace{1.5em}}c@{\hspace{2em}}c@{\hspace{1.5em}}c}
\hline
 & tree-level${} + {\cal O}(\alpha)$ & 
 ${\cal O}(\alpha^2_{\rm ferm}) + {\rm FSR}_{\as^3,\alpha\as,m_b^2\as,m_b^4}$ & \\[-.5ex]
$\MH$ & ${} + {\rm FSR}_{\alpha,\as,\as^2}$ & 
${} + {\cal O}(\alpha\as,\alpha\as^2)$ & total
\\[-.5ex] 
[GeV] & & [10$^{-4}$] & \\
\hline
\phantom{1}100 & 0.21569 & $-$1.923 & 0.21549 \\
\phantom{1}200 & 0.21570 & $-$1.919 & 0.21551 \\
\phantom{1}400 & 0.21572 & $-$1.916 & 0.21553 \\
\phantom{1}600 & 0.21573 & $-$1.918 & 0.21554 \\
1000           & 0.21574 & $-$1.927 & 0.21555 \\
\hline
\end{tabular}
\end{center}
\vspace{-2ex}
\mycaption{Results for $R_b$, as in Table~\ref{tab:rbres1}, but now with $\MW$
calculated from $G_\mu$ using the SM prediction. The other input
values are taken from Tab.~\ref{tab:input}.
\label{tab:rbres2}}
\end{table}
The most precise result including electroweak two-loop and
QCD three-loop corrections differs from the experimental value $R_b = 0.21629 \pm 0.00066$ \cite{pdg}
by about 1.2 standard deviations.

The evaluation of the complete fermionic two-loop result with the MB integrals
takes several CPU-hours for one set of parameters and thus is not suitable for
direct incorporation in SM fit programs. However, the prediction for $R_b$ can be approximated
to a good precision by a simple parametrization formula:
\begin{equation}
\begin{aligned}
R_b = R_b^0 &+ c_1 L_H + c_2 L_H^2 + c_3 L_H^4 + c_4 (\Delta_H^2 -1)
 + c_5 \Delta_\alpha \\
&+ c_6 \Delta_t + c_7 \Delta_t L_H + c_8 \Delta_{\alpha_s} + c_9
\Delta_{\alpha_s}^2 + c_{10} \Delta_Z,
\end{aligned}
\end{equation}
with
\begin{align}
L_H &= \ln\frac{M_H}{100~\mathrm{GeV}}, &
\Delta_H &= \frac{M_H}{100~\mathrm{GeV}}, &
\Delta_t &= \left(\frac{m_t}{173.2~\mathrm{GeV}}\right)^2 -1,
\nonumber \\[1ex]
\Delta_\alpha &= \frac{\Delta \alpha}{0.05900}-1, &
\Delta_{\alpha_s} &= \frac{\alpha_s(M_Z)}{0.1184}-1, &
\Delta_Z &= \frac{M_Z}{91.1876~\mathrm{GeV}} -1.
\end{align}
The numerical coefficients are determined by a fit to the full numerical result,
which includes all radiative corrections mentioned above: 
the complete ${\cal O}(\alpha)$ and fermionic ${\cal
O}(\alpha^2)$ contributions to the $Zf\bar{f}$ vertex form factors,
as well as virtual ${\cal O}(\alpha\as)$ and ${\cal O}(\alpha\as^2)$
corrections and final-state radiation of order ${\cal O}(\as^n)$, ($n=1,2,3$)
and ${\cal O}(\alpha\as)$. For the $W$-boson mass the currently most
precise result of Ref.~\cite{mwtot} is used. With these ingredients, the fit result
for the coefficients is
\begin{equation}
\begin{aligned}
R_b^0 &= 0.2154940, &
c_1 &= 1.88\times 10^{-5}, &
c_2 &= 2.0\times 10^{-6}, &
c_3 &= -6.0\times 10^{-7}, \\
c_4 &= 8.53\times 10^{-8}, &
c_5 &= 7.05\times 10^{-4}, &
c_6 &= -3.159\times 10^{-3}, &
c_7 &= 6.65\times 10^{-5}, \\
c_8 &= -1.704\times 10^{-3}, &
c_9 &= -9.30\times 10^{-4}, &
c_{10} &= 6.26\times 10^{-2}.
\end{aligned}
\end{equation}
With this parametrization the full result is approximated to better than
$10^{-6}$ for $10\gev\leq M_H \leq 1\tev$ and the other input parameters in their
$2 \sigma$ experimental error ranges.


\section{Summary}
\label{sc:sum}

Mellin-Barnes representations provide a generic framework for evaluating
multi-loop integrals numerically. They can used to systematically extract all
physical singularities of a given loop diagram. Furthermore, by using
appropriate variable mappings and contour deformations one can achieve good
numerical convergence of the Mellin-Barnes integrals for a large class of loop
diagrams, see Ref.~\cite{mbn}.

This paper reports on the application of this method to the calculation of
fermionic electroweak two-loop corrections to the effective weak mixing angle
for bottom quarks, $\sin^2 \theta_{\rm eff}^{b\bar{b}}$, and the branching ratio
of the $Z$ boson into bottom quarks, $R_b$. These two observables have been
measured to high accuracy by experiments at LEP and SLC, and they play an
important role in precision tests of the Standard Model, as well as new physics
models. Therefore one needs theoretical predictions for these quantities with an
error that is comparable or less than the experimental uncertainty, which
requires the inclusion of electroweak two-loop corrections.

In the work presented here, the numerical Mellin-Barnes method has been used for
the most difficult two-loop vertex diagrams. For the one-loop and remaining
two-loop diagrams, a technique based on tensor reduction to a set of well-known
master integrals has been employed, which is faster but less general than the
Mellin-Barnes method.

The results for $\sin^2 \theta_{\rm eff}^{b\bar{b}}$ have been compared to a 
previous calculation of the fermionic two-loop corrections with a different
method  Ref.~\cite{swbb}. Good agreement within numerical integration errors has
been found, which serves as a validation of the numerical MB method.

The two-loop result for $R_b$ presented here is new. Its numerical value 
turns out to be relatively small, which is a coincidence
since parametrically the fermionic
electroweak two-loop terms of order ${\cal O}(N_f\alpha^2) \sim 10^{-3}$
may be large due to the number $N_f$ of fermion species running in the loops.
Thus this calculation helps
to reduce the uncertainty of the theory prediction for $R_b$. For easy use by other
groups, a simple parametrization formula has been provided,
which accurately approximates the full result within integration errors.


\section*{Acknowledgements}

This work has been supported in part by the National Science Foundation under
grant no.\ PHY-0854782.



\begin{thebibliography}{99}
\frenchspacing

\bibitem{mw}
A.~Freitas, W.~Hollik, W.~Walter and G.~Weiglein,
Phys.\ Lett.\ B {\bf 495}, 338 (2000)
[Erratum-ibid.\ B {\bf 570}, 260 (2003)],\\
M.~Awramik and M.~Czakon,
Phys.\ Rev.\ Lett.\  {\bf 89}, 241801 (2002);\\
A.~Onishchenko and O.~Veretin,
Phys.\ Lett.\ B {\bf 551}, 111 (2003);\\
M.~Awramik, M.~Czakon, A.~Onishchenko and O.~Veretin,
Phys.\ Rev.\ D {\bf 68}, 053004 (2003);\\
M.~Awramik and M.~Czakon,
Phys.\ Lett.\ B {\bf 568}, 48 (2003).

\bibitem{mwlong}
A.~Freitas, W.~Hollik, W.~Walter and G.~Weiglein,
Nucl.\ Phys.\ B {\bf 632},189 (2002)
[Erratum-ibid.\ B {\bf 666}, 305 (2003)].

\bibitem{mwtot}
  M.~Awramik, M.~Czakon, A.~Freitas and G.~Weiglein,
  Phys.\ Rev.\ D {\bf 69}, 053006 (2004).

\bibitem{swlept}
  M.~Awramik, M.~Czakon, A.~Freitas, G.~Weiglein,
  Phys.\ Rev.\ Lett.\  {\bf 93}, 201805 (2004);\\
  M.~Awramik, M.~Czakon and A.~Freitas,
  Phys.\ Lett.\  B {\bf 642}, 563 (2006).

\bibitem{swlept2}
  M.~Awramik, M.~Czakon and A.~Freitas,
  JHEP {\bf 0611}, 048 (2006).

\bibitem{swlept3}
  W.~Hollik, U.~Meier and S.~Uccirati,
  Nucl.\ Phys.\ B {\bf 731}, 213 (2005);\\
  W.~Hollik, U.~Meier and S.~Uccirati,
  Nucl.\ Phys.\ B {\bf 765}, 154 (2007).

\bibitem{swbb}
  M.~Awramik, M.~Czakon, A.~Freitas and B.~A.~Kniehl,
  Nucl.\ Phys.\  B {\bf 813}, 174 (2009).

\bibitem{mt6}
J.~J.~van der Bij, K.~G.~Chetyrkin, M.~Faisst, G.~Jikia and T.~Seidensticker,
Phys.\ Lett.\ B {\bf 498}, 156 (2001);\\
M.~Faisst, J.~H.~K\"uhn, T.~Seidensticker and O.~Veretin,
Nucl.\ Phys.\ B {\bf 665}, 649 (2003).

\bibitem{radja}
  R.~Boughezal, J.~B.~Tausk and J.~J.~van der Bij,
  Nucl.\ Phys.\ B {\bf 713}, 278 (2005);\\
  R.~Boughezal, J.~B.~Tausk and J.~J.~van der Bij,
  Nucl.\ Phys.\ B {\bf 725}, 3 (2005).

\bibitem{qcd3}
L.~Avdeev, J.~Fleischer, S.~Mikhailov and O.~Tarasov,
Phys.\ Lett.\ B {\bf 336}, 560 (1994)
[Erratum-ibid.\ B {\bf 349}, 597 (1994)];\\
K.~G.~Chetyrkin, J.~H.~K\"uhn and M.~Steinhauser,
Phys.\ Lett.\ B {\bf 351}, 331 (1995);\\
K.~G.~Chetyrkin, J.~H.~K\"uhn and M.~Steinhauser,
Phys.\ Rev.\ Lett.\  {\bf 75}, 3394 (1995).

\bibitem{qcd3light}
K.~G.~Chetyrkin, J.~H.~K\"uhn and M.~Steinhauser,
Nucl.\ Phys.\ B {\bf 482}, 213 (1996).

\bibitem{qcd4}
Y.~Schr\"oder and M.~Steinhauser,
Phys.\ Lett.\ B {\bf 622}, 124 (2005);\\
K.~G.~Chetyrkin, M.~Faisst, J.~H.~K\"uhn, P.~Maierhoefer and C.~Sturm,
  Phys.\ Rev.\ Lett.\  {\bf 97}, 102003 (2006);\\
  R.~Boughezal and M.~Czakon,
  Nucl.\ Phys.\  B {\bf 755}, 221 (2006).

\bibitem{zfitter}
  A.~B.~Arbuzov {\it et al.},
  Comput.\ Phys.\ Commun.\  {\bf 174}, 728 (2006).

\bibitem{gfitter}
  H.~Fl\"acher, M.~Goebel, J.~Haller, A.~H\"ocker, K.~M\"onig and J.~Stelzer,
  Eur.\ Phys.\ J.\  C {\bf 60}, 543 (2009)
  [Erratum-ibid.\  C {\bf 71}, 1718 (2011)].

\bibitem{mb}
  C.~Anastasiou and A.~Daleo,
  JHEP {\bf 0610}, 031 (2006).

\bibitem{mb2}
  M.~Czakon,
  Comput.\ Phys.\ Commun.\  {\bf 175}, 559 (2006).

\bibitem{ambre}
  J.~Gluza, K.~Kajda, T.~Riemann and V.~Yundin,
  Eur.\ Phys.\ J.\  C {\bf 71}, 1516 (2011).

\bibitem{mbn}
  A.~Freitas and Y.~C.~Huang,
  JHEP {\bf 1004}, 074 (2010).

\bibitem{ewmt2}
  G.~Degrassi and P.~Gambino,
  Nucl.\ Phys.\  B {\bf 567}, 3 (2000).

\bibitem{ewmt4}
R.~Barbieri, M.~Beccaria, P.~Ciafaloni, G.~Curci and A.~Vicere,
Phys.\ Lett.\ B {\bf 288}, 95 (1992)
[Erratum-ibid.\ B {\bf 312}, 511 (1993)];\\
R.~Barbieri, M.~Beccaria, P.~Ciafaloni, G.~Curci and A.~Vicere,
Nucl.\ Phys.\ B {\bf 409}, 105 (1993);\\
J.~Fleischer, O.~V.~Tarasov and F.~Jegerlehner,
Phys.\ Lett.\ B {\bf 319}, 249 (1993);\\
J.~Fleischer, O.~V.~Tarasov and F.~Jegerlehner,
Phys.\ Rev.\ D {\bf 51}, 3820 (1995).

\bibitem{ibp}
K.~G.~Chetyrkin and F.~V.~Tkachov,
Nucl.\ Phys.\ B {\bf 192}, 159 (1981);\\
T.~Gehrmann and E.~Remiddi,
Nucl.\ Phys.\ B {\bf 580}, 485 (2000).

\bibitem{bauberger}
S.~Bauberger, F.~A.~Berends, M.~B\"ohm and M.~Buza,
Nucl.\ Phys.\ B {\bf 434}, 383 (1995);\\
S.~Bauberger and M.~B\"ohm,
Nucl.\ Phys.\ B {\bf 445}, 25 (1995).

\bibitem{RVA}
  K.~G.~Chetyrkin, A.~L.~Kataev and F.~V.~Tkachov,
  Phys.\ Lett.\  B {\bf 85}, 277 (1979);\\
  M.~Dine and J.~R.~Sapirstein,
  Phys.\ Rev.\ Lett.\  {\bf 43}, 668 (1979);\\
  W.~Celmaster and R.~J.~Gonsalves,
  Phys.\ Rev.\ Lett.\  {\bf 44}, 560 (1980);\\
  S.~G.~Gorishnii, A.~L.~Kataev and S.~A.~Larin,
  Phys.\ Lett.\  B {\bf 212}, 238 (1988);\\
  K.~G.~Chetyrkin and J.~H.~K\"uhn,
  Phys.\ Lett.\  B {\bf 248}, 359 (1990);\\
  S.~G.~Gorishnii, A.~L.~Kataev and S.~A.~Larin,
  Phys.\ Lett.\  B {\bf 259}, 144 (1991);\\
  L.~R.~Surguladze and M.~A.~Samuel,
  Phys.\ Rev.\ Lett.\  {\bf 66}, 560 (1991)
  [Erratum-ibid.\  {\bf 66}, 2416 (1991)];\\
  A.~L.~Kataev,
  Phys.\ Lett.\  B {\bf 287}, 209 (1992);\\
  K.~G.~Chetyrkin,
  Phys.\ Lett.\  B {\bf 307}, 169 (1993).

\bibitem{RVA2}
K.~G.~Chetyrkin, J.~K\"uhn and A.~Kwiatkowski, in
{\it Reports of the Working Group on Precision Calculations for the Z
Resonance}, eds.\ D.~Bardin, W.~Hollik and G.~Passarino, 
report CERN 95-03 (1995), pp.~175--263.

\bibitem{RVA3}
  D.~Y.~Bardin {\it et al.},
  Comput.\ Phys.\ Commun.\  {\bf 133}, 229 (2001),
sect.~3.6.

\bibitem{pdg}
  K.~Nakamura {\it et al.}  [Particle Data Group],
  J.\ Phys.\ G {\bf 37}, 075021 (2010).

\bibitem{tevewwg}
  The Tevatron Electroweak Working Group and the CDF and D\O\ 
                  Collaborations,
  arXiv:1107.5255 [hep-ex];\\
  H.~Burkhardt and B.~Pietrzyk,
  Phys.\ Rev.\  D {\bf 84}, 037502 (2011).

\bibitem{masst}
  D.~Yu.~Bardin, A.~Leike, T.~Riemann and M.~Sachwitz,
  Phys.\ Lett.\ B 206 (1988) 546.

\bibitem{qcd2}
A.~Djouadi and C.~Verzegnassi,
Phys.\ Lett.\ B {\bf 195}, 265 (1987);\\
A.~Djouadi,
Nuovo Cim.\ A {\bf 100}, 357 (1988);\\
B.~A.~Kniehl,
Nucl.\ Phys.\ B {\bf 347}, 86 (1990);\\
B.~A.~Kniehl and A.~Sirlin,
Nucl.\ Phys.\ B {\bf 371}, 141 (1992);\\
A.~Djouadi and P.~Gambino,
Phys.\ Rev.\ D {\bf 49}, 3499 (1994)
[Erratum-ibid.\ D {\bf 53}, 4111 (1996)].

\end{thebibliography}
\end{document}